\documentclass[twocolumn]{aastex631}

\usepackage{xspace}

\newcommand{\R}{\mathcal{R}}
\newcommand{\T}{\mathcal{T}}

\newcommand{\Z}{12+\rm{log}(O/H)}
\usepackage{graphicx,grffile,ulem,url,threeparttable,xcolor,soul,booktabs,natbib,amsmath,mathrsfs}
\usepackage{txfonts,rotating,longtable, needspace,subfigure,multirow,array}

\newcommand{\oii}{\text{[\ion{O}{2}]}}

\newcommand{\oiii}{\text{[\ion{O}{3}]}}

\newcommand{\nii}{\text{[\ion{N}{2}]}}

\newcommand{\ha}{\text{H$\alpha$}}

\newcommand{\hb}{\text{H$\beta$}}

\newcommand{\hii}{\text{\ion{H}{2}}}

\newcommand{\sii}{\text{[\ion{S}{2}]}}

\newcommand{\comment}[1]{}

\renewcommand{\ne}{\text{$n_{\rm e}$}}

\shorttitle{Linking Electron Density with Elevated Star Formation Activity from $z=0$ to $z=10$}
\shortauthors{Li et al.}

\begin{document}

\title{Linking Electron Density with Elevated Star Formation Activity from $z=0$ to $z=10$}

\author[0000-0003-4813-8482]{Sijia Li}
\affiliation{Department of Astronomy, Xiamen University, Xiamen, Fujian 361005, China}

\author[0000-0002-3462-4175]{Si-Yue Yu}
\affiliation{Kavli Institute for the Physics and Mathematics of the Universe (WPI), The University of Tokyo Institutes for Advanced Study, The University of Tokyo,\\ Kashiwa, Chiba 277-8583, Japan}
\affiliation{Department of Astronomy, Xiamen University, Xiamen, Fujian 361005, China}

\author[0000-0001-6947-5846]{Luis C. Ho}
\affiliation{The Kavli Institute for Astronomy and Astrophysics, Peking University, 5 Yiheyuan Road, Haidian District, Beijing, 100871, China}
\affiliation{Department of Astronomy, Peking University, 5 Yiheyuan Road, Haidian District, Beijing, 100871, China}

\author[0000-0002-0000-6977]{John D. Silverman}
\affiliation{Kavli Institute for the Physics and Mathematics of the Universe (WPI), The University of Tokyo Institutes for Advanced Study, The University of Tokyo,\\ Kashiwa, Chiba 277-8583, Japan}
\affiliation{Department of Astronomy, School of Science, The University of Tokyo, 7-3-1 Hongo, Bunkyo, Tokyo 113-0033, Japan}
\affiliation{Center for Data-Driven Discovery, Kavli IPMU (WPI), UTIAS, The University of Tokyo, Kashiwa, Chiba 277-8583, Japan}
\affiliation{Center for Astrophysical Sciences, Department of Physics \& Astronomy, Johns Hopkins University, Baltimore, MD 21218, USA}

\author[0000-0002-6593-8820]{Jing Wang}
\affiliation{The Kavli Institute for Astronomy and Astrophysics, Peking University, 5 Yiheyuan Road, Haidian District, Beijing, 100871, China}
\affiliation{Department of Astronomy, Peking University, 5 Yiheyuan Road, Haidian District, Beijing, 100871, China}

\author[0000-0003-4357-3450]{Amélie Saintonge}
\affiliation{Max-Planck-Institut für Radioastronomie, Auf dem Hügel 69, 53121 Bonn, Germany}

\author[0000-0002-9066-370X]{Niankun Yu}
\affiliation{Max-Planck-Institut für Radioastronomie, Auf dem Hügel 69, 53121 Bonn, Germany}

\author[0000-0001-7232-5355]{Qinyue Fei}
\affiliation{David A. Dunlap Department of Astronomy and Astrophysics, University of Toronto, 50 St. George Street, Toronto, Ontario, M5S 3H4, Canada}

\author[0000-0001-9044-1747]{Daichi Kashino}
\affiliation{National Astronomical Observatory of Japan, 2-21-1 Osawa, Mitaka, Tokyo 181-8588, Japan}

\author[0000-0001-5277-4882]{Hao-ran Yu}
\affiliation{Department of Astronomy, Xiamen University, Xiamen, Fujian 361005, China}

\correspondingauthor{Si-Yue Yu}
\email{si-yue.yu@ipmu.jp}

\begin{abstract}
The interstellar medium (ISM) in high-redshift galaxies exhibits significantly higher electron densities (\ne) than in the local universe. To investigate the origin of this trend, we analyze a sample of 9590 centrally star-forming galaxies with stellar masses greater than $10^9\,M_\odot$ at redshifts $0.01 < z < 0.04$, selected from the Dark Energy Spectroscopic Instrument (DESI) Data Release 1. We derive electron densities from the $\sii$\,$\lambda\lambda 6716, 6731$ doublet, measuring values of $\ne = 30$--$400\,\text{cm}^{-3}$ at $z \approx 0$. We find a tight correlation between \ne\ and the star formation rate surface density ($\Sigma_{\text{SFR}}$), which is well-described by a broken power law. Above a threshold of $\log\,(\Sigma_{\text{SFR}} / M_\odot\,\text{yr}^{-1}\,\text{kpc}^{-2}) \geq -1.46$, the relation follows $\ne = (233 \pm 13) \times \Sigma_{\text{SFR}}^{0.49 \pm 0.02}$. Below this threshold, \ne\ remains approximately at a constant value of $44 \pm 3\,\text{cm}^{-3}$. Remarkably, this relation remains consistent with measurements of galaxies at $z = 0.9$--$10.2$.  By converting the observed redshift evolution of $\Sigma_{\rm SFR}$ into \ne\ evolution through our \ne–$\Sigma_{\rm SFR}$ relation, we obtain $\ne = 40\times (1+z)^{1.4}\,\text{cm}^{-3}$, consistent with previous direct observations. The \ne-$\Sigma_{\rm SFR}$ relation arises likely because the high $\Sigma_{\rm SFR}$, fueled by dense cold gas or elevated efficiency, enhances radiative and mechanical feedback and produces dense ionized gas whose electron densities are further regulated by ambient pressure. We conclude that the redshift evolution of \ne\ primarily reflects the evolution of the cold gas density and star formation activity over cosmic time.
\end{abstract}


\section{Introduction}  \label{sec:intro}

The electron density (\ne) in the interstellar medium (ISM) is a key parameter for constraining the structure and ionization state of the \hii\ regions in galaxies \citep[e.g.,][]{Brinchmann2008, 2008Liu, Sanders2016, Ji2020, NYU2024}. 
\ne\ is commonly determined from the flux ratios of the $\oii\lambda\lambda 3726,3729$ or $\sii\lambda\lambda 6716,6731$ doublet.  Typical \hii\ regions in nearby star-forming galaxies generally have $\ne=50$--$100$\,cm$^{-3}$ \citep{1979Pagel_HIIinNGC300, 1986Dopita&Evans, Osterbrock1989, Brinchmann2008, Shirazi2014}. Further studies explored \ne\ in galaxies at $z\sim1$--$2$ and find \ne\ increases by up to an order of magnitude \citep[e.g.,][]{Lehnert2009, Masters2014, Shimakawa2015,Bian2016,Kashino2017,Sanders2016,Steidel2016,Davies2021}. With the launch of JWST, the redshift has been extended to 10, allowing recent works to reveal its redshift evolution: $\ne\propto(1+z)^p$, where $p\approx1$--$2$ \citep{isobe2023redshift,2023Reddy_2.7-6.3,Reddy2023SFR,Abdurrouf_2024,2025Sijia,2025Topping}.

What drives the redshift evolution of \ne? High-redshift galaxies are observed to be more gas-rich \citep[e.g.,][]{2010Tacconi,2010Daddi}, more actively star-forming \citep[e.g.,][]{Speagle2014},  and more compact \citep[e.g.,][]{van_der_Wel_2014}, all of which influence ionization conditions and are likely key contributors to the redshift evolution of \ne. The electron density in high-redshift galaxies has been shown to increase with star formation rate (SFR) surface density ($\Sigma_{\rm SFR}$) following a power-law relation \citep{Shimakawa2015, Jiang2019, Davies2021, Reddy2023SFR}. \citet{Davies2021} interpreted the \ne\ evolution as the result of an interplay between the high gas density (and thus high $\Sigma_{\rm SFR}$) and the subsequent effects of stellar feedback and ambient pressure.

It is worth noting that high electron densities of several hundred cm$^{-3}$ are also observed in actively star-forming, particularly starburst, regions of nearby galaxies \citep{Lehnert1996, Ho_2003, HuntHirashita2009, Lehnert2009, 2011Westmoquette, 2013Westmoquette, 2015McLeod, 2016Herrera-Camus, 2018Kakkad}. This implies that the high \ne\ at high redshift may reflect physical conditions analogous to local starbursts, driven by shared underlying mechanisms. Therefore, comparative studies of galaxies spanning a wide range of \ne\ values, at both low and high redshifts, while controlling for key galaxy properties such as $\Sigma_{\rm SFR}$, are essential to uncover the nature of its redshift evolution. Using SDSS spectra of nearby galaxies, earlier works \citep[e.g.,][]{2008Liu, Kaasinen2017,2019Kashino_and_Inoue} have correlated central \ne\ with total stellar mass or total SFR, finding qualitatively that massive, high-SFR systems exhibit higher \ne. However, the correlation between \ne\ and $\Sigma_{\rm SFR}$ in nearby galaxies and the comparison with high-redshift measurements have not yet been studied.

In this work, we use data from the spectroscopic redshift survey of the Dark Energy Spectroscopic Instrument (DESI) to examine the correlation between \ne\ and $\Sigma_{\rm SFR}$ in the central regions sampled by the DESI fiber for nearby galaxies. We then compare our low-redshift results with high-redshift results from previous works to investigate the nature of the redshift evolution of \ne. Throughout this paper, we adopt a Chabrier initial mass function (IMF) \citep{Chabrier2003} and a flat $\Lambda$CDM cosmology with $(\Omega_{\rm m}, \Omega_{\Lambda}, h) = (0.27, 0.73, 0.70)$.

\begin{figure*}
    \centering
    \includegraphics[width=1.0\linewidth]{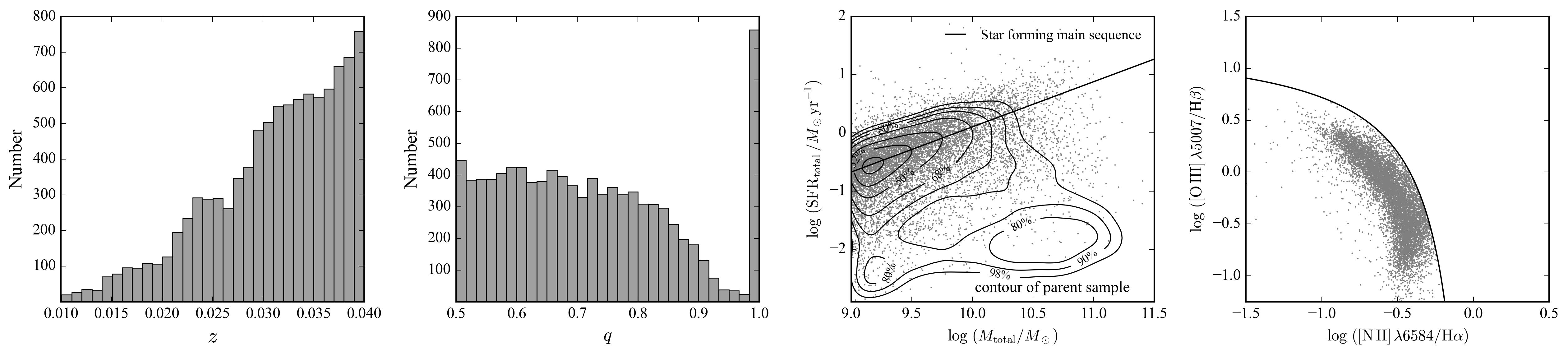}
    \caption{Basic properties of our galaxy sample. 
    From left to right, the panel present the histogram of redshift ($z$), axis ratio ($q$), the diagram of total stellar mass ($M_{\rm total}$) versus total SFR (${\rm SFR_{total}}$), and the BPT diagram of $\oiii\,\lambda5007/\hb$ versus $\nii\,\lambda6584/\ha$. { The straight line represents the star-forming main sequence for galaxies at $z\leq0.04$, and the contours indicate regions enclosing a given fraction of the galaxies in the parent sample.}
    The solid curve in the BPT diagram marks the empirical separation between star-forming galaxies and AGNs, as defined by \cite{Kauffmann2003}.}
    \label{fig:sample}
\end{figure*}

\section{Sample and data} \label{sec:data}

We use the redshift catalog from DESI Data Release 1 \citep[DR1;][]{EDR_DESI2024, DR1_DESI2025}\footnote{\url{https://data.desi.lbl.gov/public/dr1/spectro/redux/iron/zcatalog/v1/zall-pix-iron.fits}} to define our sample. Following the DR1 guidelines, we require ${\tt ZWARN} = 0$, ${\tt DELTACHI2} > 25$, and ${\tt ZCAT~PRIMARY} = 1$ to ensure robust redshift measurements. Our sample is drawn from Bright Galaxy Survey (BGS) targets classified as galaxies in the DR1 spectral analysis. The BGS galaxies have redshift $z \lesssim 0.5$. For this study, we select galaxies with $0.01 \leq z \leq 0.04$, ensuring fiber coverage of the center within $\sim$\,1\,kpc. We have verified that adopting different redshift limits does not affect our results. 
From the value-added catalog by \citet{Zou2024}\footnote{\url{https://data.desi.lbl.gov/doc/releases/dr1/vac/stellar-mass-emline/}}, we obtain effective radius ($R_e$) and axis ratio ($q=b/a$, where $a$ and $b$ is galaxy semi-major and semi-minor axis), derived from Sérsic model fitting; { in-fiber stellar masses ($M_{\rm fiber}$)}, derived from spectra fitting using {\tt StarLight} \citep{Fernandes2005}; and total stellar masses ($M_{\rm total}$) and total SFRs (${\rm SFR_{total}}$), derived from SED fitting. 
The fluxes and uncertainties of narrow emission lines (\ha, \hb, $\nii\,\lambda6584$, $\oiii\,\lambda5007$, and the $\sii\,\lambda\lambda6716, 6731$ doublet) and the Gaussian emission-line width before convolution for \sii\ doublet ($\sigma_{\rm [S\,II]}$) are taken from the version~3.0 catalog created by {\tt FastSpecfit} \citep{Moustakas2023} \footnote{\url{https://data.desi.lbl.gov/public/dr1/vac/dr1/fastspecfit/iron/v3.0/}}. In \texttt{FastSpecFit}, emission lines are modeled jointly with the stellar continuum using stellar population synthesis templates. Doublets, such as \sii\,$\lambda\lambda 6716,6731$, are fitted with Gaussians that share the same kinematics, and their flux ratio is treated as a free parameter, allowing it to serve as an electron density diagnostic.

We limit our analysis to galaxies with total stellar mass greater than $10^9\,M_\odot$, { which define the parent sample.} To properly calculate $\Sigma_{\rm SFR}$, it is necessary to correct for the inclination by deprojecting the fiber area (next section). However, this becomes ill-defined for edge-on systems, so we restrict our sample to relatively face-on galaxies with $q \geq 0.5$. Additionally, if a galaxy is much smaller than the fiber, the in-fiber $\Sigma_{\rm SFR}$ will underestimate the intrinsic value. To avoid this, we require an effective radius of at least $R_e/q \geq 0 \farcs 75$ , which corresponds to half the DESI fiber diameter ($d=1\farcs5$, corresponding to 0.3--1.2 kpc at $ 0.01\leq z\leq 0.04 $), { and 178 galaxies are excluded}. To ensure reliable emission line measurements, we require that the \ha\ and \hb\ fluxes exceed 3 times their respective uncertainties, and that the $\nii\,\lambda6584$, $\oiii\,\lambda5007$, and $\sii\,\lambda\lambda6716, 6731$ fluxes exceed their uncertainties by at least a factor of 1. We restrict our sample to star-forming galaxies, identified using the \citet{Kauffmann2003} demarcation between star-forming galaxies and AGNs. After applying all selection criteria, our final sample consists of 9590 galaxies. The distributions of $z$, $q$, the $M_{\rm total}$-${\rm SFR_{total}}$ diagram, and the Baldwin-Phillips-Terlevich (BPT) diagram \citep{1981BPT} are presented in Fig.~\ref{fig:sample}. 
{
In the $M_{\rm total}$-${\rm SFR_{total}}$ diagram, the contours indicate regions enclosing a given fraction of the galaxies in the parent sample. 
Most galaxies in our sample follow the star-forming main sequence:  
\begin{equation}
    \log\,({\rm SFR}/M_{\odot}\, {\rm yr^{-1}}) = 0.776 \times \log\,{(M_*/M_{\odot})} - 7.657,     
\end{equation}
which is calculated for galaxies at $z\leq0.04$ using the same catalog data. }
Although some galaxies fall in the green valley and a few in the quenched region, { their centers are star-forming based on the \ha\ emission.} The \ne\ can also be measured from the $\oii\,\lambda\lambda3726, 3729$ line ratio, which yields results consistent with those from the \sii\ doublet \citep{Sanders2016}. In this work, we instead use the $\sii\,\lambda\lambda6716, 6731$ lines, as the \oii\ doublet has lower signal-to-noise and its lines are closely spaced and not well resolved.

\section{Method} \label{sec:method}
Galaxy centers { may} exhibit higher molecular gas densities and SFR surface densities than their outskirts, due to the inward gas transport driven by spiral arms and bars \citep{Regan1999, Sheth2005, Kuno2007, Lin2017, Lin2020, Wang2020, Yu2022a, Yu2022b, XuYu2024, Yu2025}, despite some early-type disk galaxies presenting quenched centers \citep{Lin2020, Wang2020}. { Focusing on the central regions allows us to probe dense and actively star-forming environments, which may share a similar high level of electron densities observed at high redshift. }
Since \ne\ also tends to decline with radius \citep{Burman2025}, averaging SFR and \ne\ over entire galaxies can dilute their intrinsic correlation { unless SFR and $n_{\rm e}$ are well correlated}. By focusing on the central region probed by the DESI fiber, we mitigate such { potential} averaging effects and better trace their intrinsic connection.

We computed the in-fiber dust-corrected SFR from the extinction-corrected H$\alpha$ luminosity following the method of \citet{Kennicutt1998ARA&A}, with a conversion factor adjusted for a Chabrier IMF. First, the color excess $E(B-V)$ is derived from the Balmer decrement, by assuming extinction curve coefficients $k(\mathrm{H}\alpha) = 3.33$ and $k(\mathrm{H}\beta) = 4.60$ and intrinsic H$\alpha$/H$\beta$ ratio of 2.86.  The attenuation at H$\alpha$ is: $A_{\mathrm{H}\alpha} = k(\mathrm{H}\alpha) \times E(B-V)$, and the extinction-corrected H$\alpha$ flux is given by $F_{\mathrm{H}\alpha}^{\mathrm{corr}} = F_{\mathrm{H}\alpha} \times 10^{0.4 A_{\mathrm{H}\alpha}}$. $L_{\mathrm{H}\alpha}$ is then calculated and converted to in-fiber SFR:

\begin{equation}
\mathrm{SFR_{fiber}}/M_\odot~\mathrm{yr}^{-1} = 4.6 \times 10^{-42} \times L_{\mathrm{H}\alpha}/\mathrm{erg}~\mathrm{s}^{-1}.
\end{equation}

\noindent
By converting the fiber diameter $d$ in arcsec to a physical scale $D$ in kpc, the surface density of the SFR within the fiber ($\Sigma_{\rm SFR}$) is defined as:
\begin{equation}
    \Sigma_{\rm SFR} = \frac{\rm SFR_{fiber}}{\pi (D/2)^2q},
\end{equation}
where $q$ removes the inclination effect. 
{ The in-fiber surface density of the stellar mass ($\Sigma_*$) is defined as:
\begin{equation}
    \Sigma_* = \frac{{M_{\rm fiber}}}{\pi (D/2)^2q}.
\end{equation}
The specific SFR ${\rm sSFR}={\rm SFR_{fiber}}/M_{\rm fiber}$ is then calculated.
}

The flux ratio $\sii\,\lambda6716/\sii\,\lambda6731$ is sensitive to \ne\ over the range $10 \lesssim \ne/{\rm cm}^{-3} \lesssim 5000$, while remaining largely insensitive to temperature \citep{2006Osterbrock}. The ratio $\sii\,\lambda6716/\sii\,\lambda6731$ is converted to \ne\ using {\tt PyNeb} \citep{Luridiana2015}, assuming a fixed electron temperature of $T_{\rm e} = 10^4$\,K \citep{Perez-Montero2017}. While this temperature is commonly assumed for typical \hii\ regions, $T_{\rm e}$ depends on gas-phase metallicity and can vary between $\sim$\,5000 and 20000\,K \cite[e.g.,][]{Wink1983, Andrews2013}. Such variations affect the derived \ne\ by $\lesssim\,0.2$\,dex \citep{Copetti2000}, a result we have independently confirmed with {\tt PyNeb}.

\begin{deluxetable}{ccc}
\tablewidth{0pt}
\tablecaption{\sii\ doublet line ratio and electron density (\ne) in each bin of surface density of star formation rate ($\Sigma_{\rm SFR}$). \label{tab:description}}
\tablehead{
\colhead{$\log\,(\Sigma_{\rm SFR}/M_{\odot}{\rm yr}^{-1}{\rm kpc}^{-2})$} & \colhead{ $\frac{[{\rm S\,II}]\,\lambda6716}{[{\rm S\,II}]\,\lambda6731}$  }& \colhead{$n_{\rm e}$/$\rm cm^{-3}$} \\
\colhead{(1)} & \colhead{(2)} & \colhead{(3)}
}
\startdata
$-2.81$ & $1.401 \pm 0.019$ & $48\pm18$ \\
$-2.43$ & $1.392 \pm 0.009$ & $58\pm9$ \\
$-2.05$ & $1.409 \pm 0.005$ & $42\pm5$ \\
$-1.68$ & $1.408 \pm 0.004$ & $42\pm3$ \\
$-1.30$ & $1.395 \pm 0.003$ & $55\pm2$ \\
$-0.92$ & $1.369 \pm 0.003$ & $80\pm4$ \\
$-0.54$ & $1.336 \pm 0.004$ & $116\pm4$ \\
$-0.17$ & $1.253 \pm 0.006$ & $218\pm7$ \\
0.21    & $1.219 \pm 0.011$ & $266\pm16$ \\
0.59    & $1.157 \pm 0.029$ & $365\pm50$ \\
\enddata
\tablecomments{Col. (1): star formation rate surface density; Col. (2): median line ratio; Col. (3): electron density derived from the median line ratio.}
\end{deluxetable}

\section{Results} \label{sec:results}

\subsection{The \sii\ line ratio, \ne, and $\Sigma_{\rm SFR}$ for our low-redshift DESI sample}

\begin{figure*}
    \centering
    \includegraphics[width=0.9\linewidth]{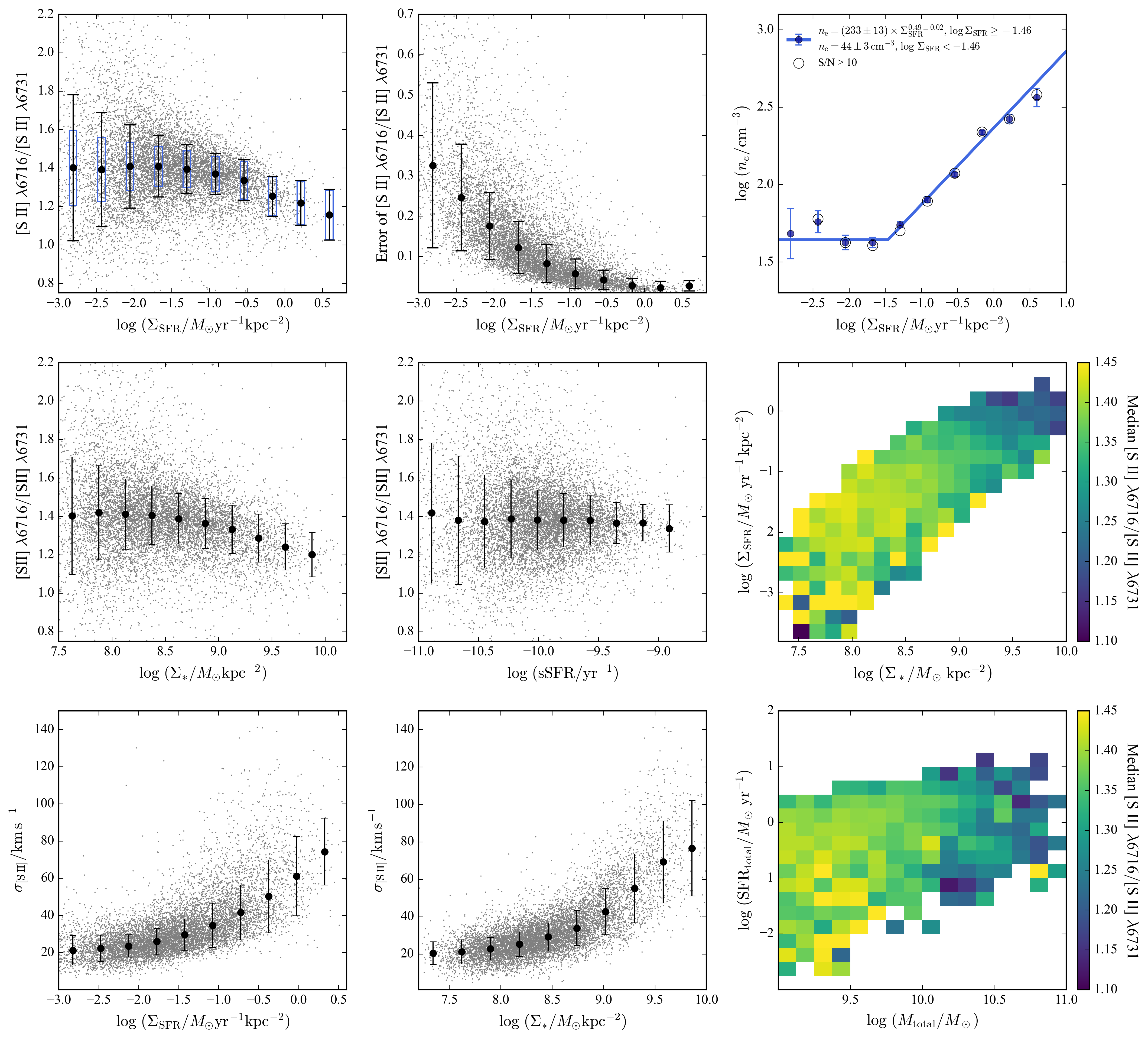}
    \caption{Relations between the \sii\ doublet ratio (\sii\,$\lambda6716/\lambda6731$), electron density (\ne), and line width ($\sigma_{\rm [S\,II]}$) with various galaxy properties. Gray dots represent individual galaxies, while black points with error bars indicate the median and scatter in each bin. In the \ne–$\Sigma_{\rm SFR}$ panel, black points show the median \ne\ derived from the median line ratio, with error bars denoting the uncertainties of the median; the best-fit broken power-law is shown in blue. { Circles in this panel mark the median values for galaxies with $S/N>10$ in all emission lines used. In the $M_{\rm total}$–${\rm SFR_{total}}$ and $\Sigma_*$–$\Sigma_{\rm SFR}$ panels, the color scale encodes the median line ratio in each bin, and bins with fewer than 5 galaxies are omitted. }
    }
    \label{fig:line_ratio}
\end{figure*}

We first investigate the dependence of the \sii\ doublet line ratio and \ne\ on the $\Sigma_{\rm SFR}$. In the top-left panel of Fig.~\ref{fig:line_ratio}, we plot the line ratio of individual galaxies as gray dots. The data show a decreasing trend of the line ratio with increasing $\Sigma_{\rm SFR}$. We divide the sample into 10 bins of $\log\Sigma_{\rm SFR}$, compute the median line ratio in each bin, and plot the median values as black points with error bars indicating the scatter. The line ratio remains largely unchanged for low $\Sigma_{\rm SFR}$, but then decreases from 1.369 at $\Sigma_{\rm SFR} = 10^{-0.92}\,M_{\odot}{\rm yr}^{-1}{\rm kpc}^{-2}$ to 1.157 at $\Sigma_{\rm SFR} = 10^{0.59}\,M_{\odot}{\rm yr}^{-1}{\rm kpc}^{-2}$. The intrinsic scatter is calculated as the quadratic difference between the total observed scatter and the median measurement error of the \sii\ ratio in each bin and is displayed as the blue boxes. The measurement error of \sii\ ratio is shown in the top-middle panel, showing the error and scatter are increasing toward lower $\log\Sigma_{\rm SFR}$, due to the fainter \sii\ doublet in lower $\log\Sigma_{\rm SFR}$ regions. The intrinsic scatter of the \sii\ line ratio in each bin does not significantly vary with $\log\Sigma_{\rm SFR}$.

As shown in Fig.~\ref{fig:line_ratio}, the typical error in the \sii\ line ratio is $\sim\,0.32$ for an individual galaxy at the lowest $\Sigma_{\rm SFR}$ bin.  This results in some measurements exceeding the theoretical upper limit of the \sii\ line ratio, where the relation with \ne\ breaks down and no reliable electron density can be derived and some with too low ratios corresponding to artificially high densities of $\sim\,1000$\,cm$^{-3}$. 
These effects arise from low-$S/N$ spectra. To overcome this issue, we compute the median line ratio in each bin and estimate its uncertainty as the scatter divided by the square root of the number of galaxies; the resulting median line ratio and its uncertainty are then converted to median \ne\ and its corresponding error. The resulting { median} \ne\ values and errors { in each bin} are listed in Table~\ref{tab:description} and shown in the top-right panel of Fig.~\ref{fig:line_ratio}. In the \ne-$\Sigma_{\rm SFR}$ relation, there exists a clear trend: at low $\Sigma_{\rm SFR}$, \ne\ remains approximately constant, while at high $\Sigma_{\rm SFR}$, $\log\ne$ increases linearly with $\log\Sigma_{\rm SFR}$. To quantify this behavior, we fit a broken power-law function and obtain:

\begin{equation}\label{bestfit}
n_{\rm e} =
\left\{
\begin{array}{ll}
(233 \pm 13) \,\times\,\Sigma_{\rm SFR}^{0.49\pm0.02}, & \text{for } \log\,\Sigma_{\rm SFR} \geq -1.46 \\
44\pm3, & \text{for } \log\,\Sigma_{\rm SFR} < -1.46
\end{array}
\right.
\end{equation}

\noindent
where \ne\ and $\Sigma_{\rm SFR}$ is in cm$^{-3}$ and $M_{\odot}\,{\rm yr}^{-1}\,{\rm kpc}^{-2}$, respectively. The best-fit \ne-$\Sigma_{\rm SFR}$ relation is plotted as a blue solid line in Fig.~\ref{fig:line_ratio}.

{ We also calculate the median \ne\ for galaxies with both \ha\ and \sii\ lines detected at $S/N>10$, marked by open circles. The overall trend remains unchanged, except that the lowest $\Sigma_{\rm SFR}$ bin no longer has sufficient data. 
We note that since the \ha\ emission flux correlates with the \sii\ doublet flux, higher $\Sigma_{\rm SFR}$ values are associated with stronger \ha\ emission and hence stronger \sii\ doublet lines. Therefore, the $S/N$ threshold primarily affects the low-$\Sigma_{\rm SFR}$ bin. Moreover, there exists a weak trend that the \sii\ doublet ratio decreases with increasing \sii\,$\lambda$6731 flux, owing to their common connection with the \ha\ line. Excluding weak \sii\ doublets thus prevents us from capturing \ne\ in the low-$\Sigma_{\rm SFR}$ bin, and we therefore adopt Eq.~(\ref{bestfit}) throughout this work. Nevertheless, we provide the best-fit relation for emission lines with $S/N>10$:
\begin{equation}\label{SNR10}
n_{\rm e} =
\left\{
\begin{array}{ll}
(235 \pm 17) \,\times\,\Sigma_{\rm SFR}^{0.52\pm0.02}, & \text{for } \log\,\Sigma_{\rm SFR} \geq -1.46 \\
41\pm4, & \text{for } \log\,\Sigma_{\rm SFR} < -1.46
\end{array}
\right.
\end{equation}
\noindent
which is fully consistent with Eq.~(\ref{bestfit}).
}
High-redshift studies, limited by smaller samples, often rely on stacked spectra; with our statistically large dataset, the median-based method is sufficient. For verification, we also generated stacked spectra in each bin and confirmed that they yield consistent electron densities.

{ 
The middle-left panel of Fig.~\ref{fig:line_ratio} shows the \sii\ line ratio as a function of $\Sigma_*$, exhibiting a decreasing trend similar to that with $\Sigma_{\rm SFR}$. The middle-center panel presents the line ratio versus sSFR, where no clear dependence is observed. The middle-right panel displays the binned distribution in the $\Sigma_*$–$\Sigma_{\rm SFR}$ plane, with color indicating the average line ratio: the ratio decreases with both increasing $\Sigma_*$ and $\Sigma_{\rm SFR}$, and there is no clear rise of $n_{\rm e}$ with higher $\Sigma_{\rm SFR}$ at fixed $\Sigma_*$. This behavior is consistent with the star-forming main sequence, which emerges because instantaneous SFR (or $\Sigma_{\rm SFR}$) is governed by the available gas supply, which is set by cosmological inflow, gas depletion, and feedback, while $\Sigma_*$ reflects the time-integrated star formation \citep{Lilly2013, Speagle2014}; therefore $\Sigma_*$ and $\Sigma_{\rm SFR}$ are tightly coupled through gas-regulation processes. Consequently, from our data alone we cannot distinguish whether $\Sigma_{\rm SFR}$ or $\Sigma_*$ is the more fundamental driver of high electron densities. Physically, however, we argue that $\Sigma_{\rm SFR}$ is more directly connected to $n_{\rm e}$ because it immediately traces the photoionization and mechanical energy input that produce and shape the ionized gas.
}

The bottom-right panel shows the variation of the \sii\ line ratio in the $M_{\rm total}$–${\rm SFR_{total}}$ plane. Galaxies with higher total stellar mass exhibit lower line ratios, corresponding to higher central \ne. At total stellar mass less than $10^{10}M_\odot$, galaxies with higher ${\rm SFR_{total}}$ for a given stellar mass exhibit lower line ratios and therefore higher central \ne.  These trends are consistent with the results in \cite{2019Kashino_and_Inoue}.

Emission-line width provides a key diagnostic of the kinematic state of the ionized gas, reflecting both intrinsic turbulence and unresolved motions such as rotational shear \citep[e.g.,][]{2009Ho}. We thus examine dependence of the line width of the \sii\ doublet ($\sigma_{\rm [S\,II]}$) on two parameters, $\Sigma_{\rm SFR}$ and $\Sigma_*$, in the bottom panels of Fig.~\ref{fig:line_ratio}. { Both $\Sigma_{\rm SFR}$ and $\Sigma_*$ shows close connection to the increase of $\sigma_{\rm [S\,II]}$.} The observed $\sigma_{\rm [S\,II]}$ in DESI fiber spectra reflects a combination of the intrinsic gas velocity dispersion and unresolved velocity gradients within the fiber aperture. { The observed increase of $\sigma_{\rm [S\,II]}$ with $\Sigma_{\rm SFR}$ can be only partially attributed to turbulence driven by stellar feedback \citep[e.g.,][]{Ostriker2011, Krumholz2018}. The line widths in the highest $\Sigma_{\rm SFR}$ bins reach up to $\sim\,80$\,km\,s$^{-1}$, significantly greater than the $\sim$\,10--20\,km\,s$^{-1}$ expected from kinetic energy injection by stellar feedback \citep{Swinbank2012, Kohandel2020}. This implies that an additional energy source, most likely the deep gravitational potential associated with the high $\Sigma_*$, is required to sustain the observed turbulence.
}

\subsection{Comparison with high-redshift results}

\begin{figure*}
    \centering
    \includegraphics[width=0.8\linewidth]{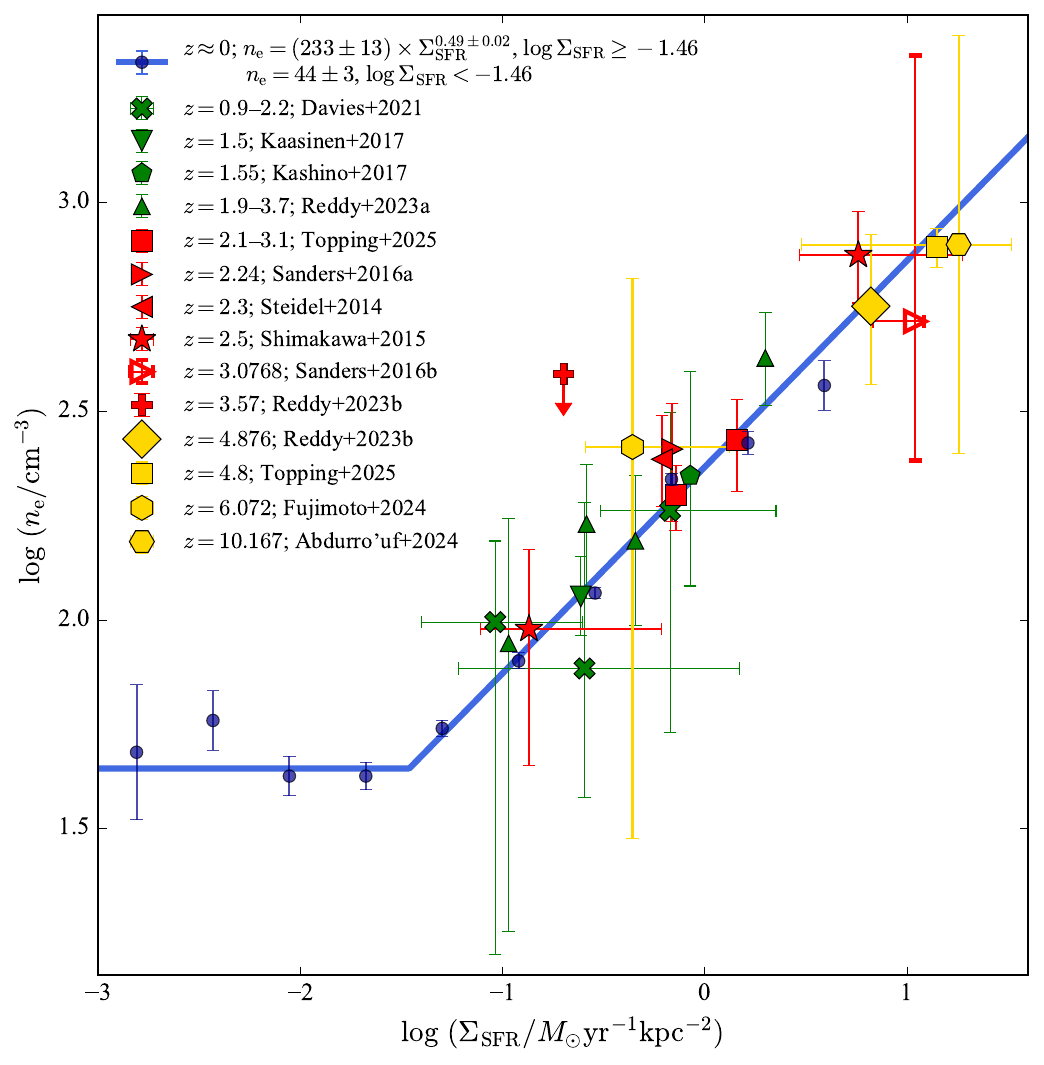}
    \caption{Relationship between electron density (\ne) and star formation rate surface density ($\Sigma_{\rm SFR}$). Our low-redshift measurements and the corresponding best-fit relation are shown as blue points and a blue curve. High-redshift measurements ($z=0.9\textendash10.2$) from previous studies are over-plotted \citep{Steidel2014,Shimakawa2015,Sanders2016,Sanders2016b,Kaasinen2017,Kashino2017,Davies2021,2023Reddy_2.7-6.3,Reddy2023SFR, Abdurrouf_2024, Fujimoto2024NA, 2025Topping}. 
    }
    \label{fig:ne}
\end{figure*}

We compare our low-redshift results with measurements of high-redshift galaxies in Fig.~\ref{fig:ne}. The flattening of \ne\ at 44\,cm$^{-3}$ for low $\Sigma_{\rm SFR}$ is consistent with \citet{Davies2021}, who measured $\sim\,31\,\mathrm{cm}^{-3}$ from the \sii\ doublet in the nearby SAMI sample across a similar low $\Sigma_{\rm SFR}$ range (see their Figure~6). The derived best-fit power-law relation is consistent with \citet{Reddy2023SFR}, who found $\ne \propto \Sigma_{\rm SFR}^{0.51 \pm 0.09}$ for a sample of 317 galaxies in the redshift range $z = 1.9$–$3.7$. \cite{Shimakawa2015} report a slightly steeper relation with a power-law index of $\approx\,$0.61 at $z=2.5$, which might have large uncertainty due to their small sample size of only 14 galaxies.  We compile measurements of $n_{\rm e}$ for galaxies with available $\Sigma_{\rm SFR}$ from previous studies covering the redshift range $0.9 \leq z \leq 10.2$ \citep{Steidel2014,Shimakawa2015,Sanders2016,Sanders2016b,Kaasinen2017,Kashino2017,Davies2021,2023Reddy_2.7-6.3,Reddy2023SFR, Abdurrouf_2024, Fujimoto2024NA, 2025Topping} and plot them in Fig.~\ref{fig:ne}. To ensure consistency, we restrict to star-forming galaxies without AGN and to electron densities derived from the \sii\ or \oii\ doublets, which provide consistent estimates \citep{Sanders2016b}. Reported \ne\ and $\Sigma_{\rm SFR}$ values in these works are used directly where available. For the samples of \citet{Kaasinen2017}, \citet{Kashino2017}, \citet{Sanders2016}, and \citet{Steidel2014}, $\Sigma_{\rm SFR}$ values are taken from \citet{Davies2021}. For \citet{2023Reddy_2.7-6.3}, we convert the \sii\ ratio at $z\approx3.57$ to \ne\ and derive a 2$\sigma$ upper limit at $z=4.876$ where the ratio exceeds the theoretical limit. \citet{2025Topping} report \ne\ and $\Sigma_{\rm SFR}$ for 43 galaxies, finding only a weak correlation between \ne\ and $\Sigma_{\rm SFR}$, perhaps due to the low $S/N$ of individual spectra; we adopt their redshift bins but compute inverse-variance–weighted mean \ne\ to mitigate this limitation. For the primordial rotating disk at $z=6.072$ studied by \citet{Fujimoto2024NA}, we calculate $\Sigma_{\rm SFR}$ from their reported SFR and the effective radius in \citet{Fujimoto2021}. The \ne\ and $\Sigma_{\rm SFR}$ of the brightness component of the lensed galaxy pair MACS0647–JD are taken from \citet{Abdurrouf_2024} and \citet{Hsiao2023}, respectively. Data at $0.9 \leq z \leq 2$, $2 < z \leq 4$, and $4 < z \leq 10.2$ are plotted in green, red, and gold, respectively.

Despite the wide redshift range ($0.9 \leq z \leq 10.2$), the high-redshift measurements agree well with our low-redshift \ne–$\Sigma_{\rm SFR}$ relation. The scatter around the relation is only 0.1\,dex. The lack of low-$\Sigma_{\rm SFR}$ points at high redshift is likely due to selection effects, as such faint galaxies are difficult to detect.

\begin{figure*}
    \centering
    \includegraphics[width=0.7\linewidth]{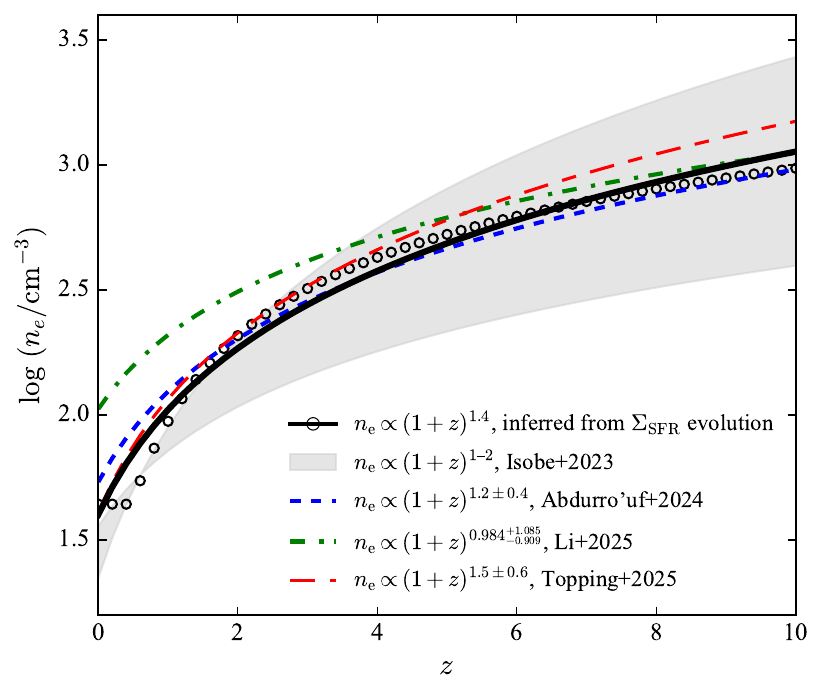}
    \caption{
    Redshift evolution of electron density $n_{\rm e}$. 
    { The open circles mark the \ne\ evolution, derived from our \ne-$\Sigma_{\rm SFR}$ relation for a fixed stellar mass of $5\times10^{10}\,M_\odot$ combined with the redshift evolution of $\Sigma_{\rm SFR}$. The black curve represents the best-fit function, $\ne = 40\times(1+z)^{1.4}\,{\rm cm}^{-3}$.} The gray shaded area shows the range estimated by \citet{isobe2023redshift}, while the dotted, dashed, and dash-dotted lines indicate trends from \citet{Abdurrouf_2024}, \citet{2025Topping}, and \citet{2025Sijia}, respectively.
    } 
    \label{fig:ne_z}
\end{figure*}

\section{Implications for the Redshift Evolution of \ne} \label{sec:implication}

The tight relation in Fig.~\ref{fig:ne} between \ne\ and $\Sigma_{\rm SFR}$ over $z=0\textendash10$ indicates that electron density in \hii\ regions is fundamentally linked to star formation activity in a manner that has remained largely invariant across cosmic time. Previous studies have primarily reported or inferred the rising trend between these two quantities \citep{Brinchmann2008, 2008Liu, 2016Masters, Bian2016, Kaasinen2017, Shimakawa2015, 2019Kashino_and_Inoue, Jiang2019, Davies2021, Reddy2023SFR}. { A direct and statistically significant power-law relation between \ne\ and $\Sigma_{\rm SFR}$ was initially established by \citet{Shimakawa2015} from 14 \ha\ emitters at $z=2.5$, and later confirmed by further investigations \citep{Jiang2019, Davies2021, Reddy2023SFR}. The observed \ne–$\Sigma_{\rm SFR}$ relation, including both the power-law regime and the low-$\Sigma_{\rm SFR}$ flattening, arises from a complex interplay among cold gas, star formation, stellar feedback, and mid-plane gravity, as discussed below.
}

{
In cold gaseous disks, giant molecular clouds collapse under self-gravity, fragment into dense cores, and form stars. Young massive stars then emit large numbers of ionizing photons that heat and ionize the surrounding gas, producing free electrons in the ISM \citep{CharlotLonghetti2001, Brinchmann2008, Ostriker2011, Kim2013}. Star formation follows the Kennicutt–Schmidt law, with higher rates in regions of denser cold gas \citep{Kennicutt1998ARA&A, Bigiel2008}. Spatially resolved observations further show that the SFR scales with the mass of gravitationally bound gas clouds \citep[e.g.,][]{Gao2004, Lada2012, Jiao2025a, Jiao2025b}, and that the mass density of a molecular cloud increases with the H$\alpha$ emission of its associated \hii\ region \citep{Zakardjian2023}. Since the ionizing photon rate increases with SFR \citep[e.g.,][]{Robertson2013, Reddy2022}, denser gas clouds naturally give rise to \hii\ regions with higher \ne, providing one potential mechanism for the observed $n_{\rm e}$–$\Sigma_{\rm SFR}$ relation.}

{
The electron density is also regulated by stellar feedback. Energy and momentum injected through stellar winds and supernovae drive the expansion of \hii\ regions \citep{Weaver1977, Oey1997, Oey1998}, which continues until internal and external pressures are balanced. This process lowers the gas density until pressure balance is reached. However, its effect is weaker than the enhancement from ionizing photons, resulting in higher \ne\ at larger $\Sigma_{\rm SFR}$.}

{
On larger scales, galaxy properties further regulate \ne. The $\Sigma_*$ and molecular disk thickness ($h_{\rm H_2}$) shape the mid-plane potential and gas density. More massive disks are denser \citep{KormendyBender2012}, and higher $\Sigma_*$ corresponds to deeper gravitational wells, which shorten the Jeans length of molecular clouds \citep{Meidt2023} and may raise \ne. Likewise, for fixed molecular surface density, thinner disks imply higher mid-plane densities ($\Sigma_{\rm H_2} \propto n_{\rm H_2}\,h_{\rm H_2}$), again favoring elevated \ne.}

{
These processes evolve with cosmic time. Compared to low-redshift systems, high-redshift galaxies are more gas-rich \citep[e.g.,][]{2010Tacconi,2010Daddi, Narayanan2012}, more compact \citep[e.g.,][]{van_der_Wel_2014, Allen2025}, and host denser disks \citep[e.g.,][]{Whitaker2017}. They also exhibit higher star formation efficiencies, with elevated $\Sigma_{\rm SFR}$ at fixed cold gas density \citep{Tacconi2013, Tacconi2020, Pallottini2022, Vallini2024}, producing more electrons while simultaneously experiencing stronger stellar feedback. There is likely a self-regulated equilibrium among star formation, stellar feedback, ionization–recombination balance, and ISM dynamics, that leads to the power-law $n_{\rm e}$–$\Sigma_{\rm SFR}$ relation consistent across redshift $z\approx0\textendash10$.}

{ 
As free electrons in the ISM are primarily produced by the ionization of cold gas from young massive stars, $n_{\rm e}$ serves as a rough tracer of the cold gas density in the multiphase ISM. Low $n_{\rm e}$ generally implies low cold gas density. Once both $n_{\rm e}$ and, correspondingly, the cold gas density fall below a critical threshold, star formation is suppressed \citep{Kennicutt1989, Schaye2004, Krumholz2009}. Galaxies in this low-density regime produce little \ha\ emission and are excluded from our sample. Nevertheless, the electron density does not vanish even at very low star formation rates, as ionization by hot, low-mass evolved stars still contributes at a minor level \citep{Fernandes2011}. At the same time, the \sii\ doublet ratio becomes insensitive to density at $\ne \lesssim 20,{\rm cm^{-3}}$. Because the \sii$\lambda$6731 line is weaker and more uncertain than \sii$\lambda$6716, measurement uncertainties tend to bias the observed ratios upward, slightly inflating the estimated median density. However, these uncertainties alone cannot fully account for the flattening feature, which remains even when adopting $S/N>10$ (Eq.~(\ref{SNR10})). The combination of these effects likely explains the flattening of the $n_{\rm e}$ trend toward low $\Sigma_{\rm SFR}$ in Fig.~\ref{fig:ne}.
}

Recently, several studies have used JWST data to constrain the redshift evolution \ne, finding a power-law relation of the form $\ne \propto (1+z)^p$ with $p \approx 1$--2 \citep{isobe2023redshift, Abdurrouf_2024, 2025Sijia, 2025Topping}.  The \ne-$\Sigma_{\rm SFR}$ enables us to use the evolution of $\Sigma_{\rm SFR}$ to infer the redshift evolution of \ne.  The increase of $\Sigma_{\rm SFR}$ with redshift { is a result of higher gas densities and higher star formation efficiency in galaxies at earlier epochs \citep[e.g.,][]{2010Tacconi,2010Daddi, Tacconi2013} and is reflected in elevated SFRs \citep[e.g.,][]{Speagle2014, Popesso2023} and more compact galaxy sizes at fixed stellar mass \citep[e.g.,][]{van_der_Wel_2014, Allen2025}.} To investigate whether the observed increase in \ne\ with redshift can be explained by $\Sigma_{\rm SFR}$ evolution, we derive the expected evolution of \ne\ from the SFR and size evolution of galaxies. For simplicity, and as our goal is a qualitative comparison, we assume a fixed stellar mass of $5\times10^{10}\,M_{\odot}$. 
{ 
The SFR evolution at this mass is taken from the star-forming main sequence parameterization by \cite{Popesso2023}: 
\begin{equation}
    \log\,({\rm SFR}/M_{\odot}\,{\rm yr}^{-1}) = -0.164\times (t/{\rm Gyr}) + 2.351,
\end{equation}
\noindent
where $t$ is the Universe age. For galaxy size, we adopt the redshift evolution for a given stellar mass of $5\times10^{10}\,M_{\odot}$ from \cite{Allen2025}:
\begin{equation}
    \log\,(R_e/{\rm kpc}) = -0.807\times \log\,(1+z) + 0.947.
\end{equation}
From these, we compute the evolution of $\Sigma_{\rm SFR}$, and use our \ne-$\Sigma_{\rm SFR}$ relation to derive the \ne\ as a function of redshift, shown as open circles in Fig.~\ref{fig:ne_z}. The curve fitting yields
\begin{equation}
    \ne = 40\times(1+z)^{1.4}\,{\rm cm}^{-3}, 
\end{equation}
which is shown as the black solid curve. 
}
For comparison, Fig.~\ref{fig:ne_z} also shows the \ne\ evolution from \cite{isobe2023redshift}, \cite{Abdurrouf_2024}, \cite{2025Sijia}, and \cite{2025Topping} as the gray shaded region and blue, green, and red dotted lines, respectively. The evolution we derive is qualitatively consistent with these direct observational constraints, implying that the redshift evolution of \ne\ is likely driven by the evolution of cold gas density and star formation activity, which are physically coupled through the Kennicutt-Schmidt law.

\section{Summary} \label{sec:conclusion}

To examine the link between electron density ($n_{\rm e}$) in the interstellar medium and the star formation rate surface density ($\Sigma_{\rm SFR}$), we use spectroscopic data from the DESI survey. Our sample comprises 9590 galaxies with total stellar mass greater than $10^{9}\,M_{\odot}$, axis ratio $q > 0.5$, effective radius $R_{\rm e}/q > 0\farcs75$ (half the DESI fiber diameter), and reliable flux measurements of \ha, \hb, $\nii\,\lambda6584$, $\oiii\,\lambda5007$, and the $\sii\,\lambda\lambda6716, 6731$ doublet. We divide the sample into 10 bins of $\log \Sigma_{\rm SFR}$, measure the median line width of $\sii$ doublet, $\sii$ doublet ratio, and derive $n_{\rm e}$.

We find that the line width of \sii\ doublet increases systematically with $\Sigma_{\rm SFR}$ and $\Sigma_*$, as stellar feedback injects turbulence into the ISM and the higher gaseous { and stellar} surface densities demand stronger turbulent support \citep{Ostriker2011,Krumholz2018}, both contributing to enhanced gas velocity dispersion.

We show that, at $\Sigma_{\rm SFR} \leq 10^{-1.46}\,M_{\odot}{\rm yr}^{-1}{\rm kpc}^{-2}$, $n_{\rm e}$ remains nearly constant at $44 \pm 3$ cm$^{-3}$. At higher $\Sigma_{\rm SFR}$, $n_{\rm e}$ rises following a power law: $n_{\rm e} = (233 \pm 13)\times\Sigma_{\rm SFR}^{0.49 \pm 0.02}$, with $n_{\rm e}$ in cm$^{-3}$ and $\Sigma_{\rm SFR}$ in $M_{\odot}{\rm yr}^{-1}{\rm kpc}^{-2}$. Moreover, our derived \ne-$\Sigma_{\rm SFR}$ relation is consistent with that reported by \cite{Reddy2023SFR} for galaxies at $z=1.9\textendash3.7$; literature measurements of \ne\ and $\Sigma_{\rm SFR}$ at $0.9 \leq z \leq 10.2$ agree with our relation. { The power-law part of the $n_{\rm e}$–$\Sigma_{\rm SFR}$ relation can be interpreted as a outcome of coupled processes in the ISM. Higher $\Sigma_{\rm SFR}$, driven by denser cold gas, enhances the injection of energy and momentum from stellar feedback, producing denser ionized regions. The resulting electron densities are further regulated by the balance between internal and ambient gas pressure and by the confining mid-plane gravity. These interconnected processes, that are star formation, stellar feedback, ionization, recombination, and ISM dynamics, collectively maintain a self-regulated equilibrium, which likely explains the near-invariance of the $n_{\rm e}$–$\Sigma_{\rm SFR}$ relation over cosmic time. The flattening of $n_{\rm e}$ at low $\Sigma_{\rm SFR}$ is plausibly caused by a critical cold gas density threshold for star formation, limitations in measuring weak \sii\ line ratios, and minor ionization contributions from hot evolved stars.}

Finally, combining our derived \ne-$\Sigma_{\rm SFR}$ relation with the observed redshift evolution of $\Sigma_{\rm SFR}$ yields a predicted scaling of $\ne = 40\times (1+z)^{1.4}\,{\rm cm}^{-3}$, consistent with previous direct observational constraints \citep{isobe2023redshift, Abdurrouf_2024, 2025Sijia, 2025Topping}. This trend indicates that the higher electron densities observed in high-redshift galaxies naturally arise from enhanced gas surface densities and more intense star formation. Our findings may also offer key insights into the higher ionization parameters characteristic of high-redshift galaxies, providing constraints on the physical conditions of star-forming gas across cosmic time.

\begin{acknowledgements}
{ We thank the referee for the insightful feedback.}
This work was supported by the National Science Foundation of China (12233001, 12173030), the Ministry of Science and Technology of the People's Republic of China (2022YFA1602902), and the China Manned Space Project (CMS-CSST-2025-A08, CMS-CSST-2025-A09). SYU acknowledges support from the UTokyo Global Activity Support Program for Young Researchers. We thank Jiaxi Yu for a fruitful discussion. This research used data obtained with the Dark Energy Spectroscopic Instrument (DESI). DESI construction and operations is managed by the Lawrence Berkeley National Laboratory. This material is based upon work supported by the U.S. Department of Energy, Office of Science, Office of High-Energy Physics, under Contract No. DE–AC02–05CH11231, and by the National Energy Research Scientific Computing Center, a DOE Office of Science User Facility under the same contract. Additional support for DESI was provided by the U.S. National Science Foundation (NSF), Division of Astronomical Sciences under Contract No. AST-0950945 to the NSF’s National Optical-Infrared Astronomy Research Laboratory; the Science and Technology Facilities Council of the United Kingdom; the Gordon and Betty Moore Foundation; the Heising-Simons Foundation; the French Alternative Energies and Atomic Energy Commission (CEA); the National Council of Humanities, Science and Technology of Mexico (CONAHCYT); the Ministry of Science and Innovation of Spain (MICINN), and by the DESI Member Institutions: www.desi.lbl.gov/collaborating-institutions. The DESI collaboration is honored to be permitted to conduct scientific research on I’oligam Du’ag (Kitt Peak), a mountain with particular significance to the Tohono O’odham Nation. Any opinions, findings, and conclusions or recommendations expressed in this material are those of the author(s) and do not necessarily reflect the views of the U.S. National Science Foundation, the U.S. Department of Energy, or any of the listed funding agencies.
\end{acknowledgements}

\bibliographystyle{aasjournal}

\end{document}